\documentclass[aps,pra,superscriptaddress,amsmath,amssymb,preprintnumbers
,showpacs,twocolumn]{revtex4}
\usepackage{graphicx,color}
\usepackage{amssymb}
\usepackage{amsmath}
\usepackage{bm}

\newcommand{\Tr}{\mathop{\rm Tr}\nolimits}

\definecolor{dgreen}{rgb}{0,0.5,0}

\definecolor{delete}{cmyk}{0.5,0,0,0}
\definecolor{deletey}{cmyk}{0,0.5,0,0}

\begin{document}


\title{A measurement scheme for purity based on two two-body gates
}


\author{H. Nakazato}\author{T. Tanaka}
\affiliation{Department of Physics, Waseda University, Tokyo 169-8555, Japan}

\author{K. Yuasa}
\affiliation{Waseda Institute for Advanced Study, Waseda University, Tokyo 169-8050, Japan}

\author{G. Florio}
\affiliation{Centro Studi e Ricerche  ``Enrico Fermi'', Piazza del Viminale 1, I-00184 Roma, Italy}
\affiliation{Dipartimento di Fisica and MECENAS, Universit\`a di Bari, I-70126 Bari, Italy}
\affiliation{INFN, Sezione di Bari, I-70126 Bari, Italy}

\author{S. Pascazio}
\affiliation{Dipartimento di Fisica and MECENAS, Universit\`a di Bari, I-70126 Bari, Italy}
\affiliation{INFN, Sezione di Bari, I-70126 Bari, Italy}


\begin{abstract}
A scheme for measuring the purity of a quantum system with a finite number of levels is presented. The method makes use of two $\sqrt{\text{SWAP}}$ gates and only hinges on measurements performed on a reference system, prepared in a certain pure state and coupled with the target system. Neither tomographic methods, with the complete reconstruction of the state, nor interferometric setups is needed.
\end{abstract}


\maketitle


\section{Introduction}
\label{sec:Intro}

Purity is a good measure of the coherence of a quantum system. 
It is unity for a system in a pure state (complete quantum coherence), while it reduces to $1/N$ for an $N$-level system in a completely mixed state (no quantum coherence).
Purity has played a central role in the discussion of the quantum measurement problem \cite{WZ83}, in the sense that the occurrence of decoherence, considered to be responsible for the transition from a quantum to a ``classical" (mixed) state by the action of measurement, is one of the main issues to be resolved within the framework of quantum theory \cite{NPN97}.
Furthermore, its importance is well recognized in the field of quantum information, communication and computation, for the purity of a subsystem is an important measure of the entanglement of the total bipartite system \cite{entp}:
the higher the entanglement between the two subsystems, the lower the purity of the reduced density matrix of the subsystems, and \emph{vice versa}. 

Although purity is an important physical notion, it is not so simple to devise an experimental procedure to directly measure it.
Remember that the quantum mechanical expectation value of an operator ${\cal O}$ in the state $\rho$ is expressed as $\Tr\{{\cal O}\rho\}$.
This is a {\it linear\/} functional of $\rho$, while purity $\Pi\equiv\Tr\{\rho^2\}$ is a {\it quadratic\/} functional. 
Purity is usually calculated only after the state $\rho$ has been reconstructed by some tomographic methods \cite{NC00}.
There are, however, interesting proposals \cite{F02,ekertetal,B04} to measure linear and nonlinear functionals of the density matrix, including purity $\Tr\{\rho^2\}$, without resorting to any tomographic methods.
One assumes that copies of the quantum system under scrutiny (target system) are available, so that the state $\rho\otimes\rho\otimes\cdots$ can be fed to an interferometric setup  in a quantum network. In particular, to determine the purity $\Tr\{\rho^2\}$ of the target system, one requires a controlled-SWAP operation and an ancillary two-level quantum system (qubit).

In this paper we present an alternative scheme for measuring the purity of a quantum system with a finite number of levels [typically a two-level system (qubit) or a three-level system (qutrit)], without resorting either to the state tomography of the density operator or to interferometry in quantum networks. 
In this way, a relatively small number of (different) measurements are required \cite{LL11}.
In particular, our scheme consists of two simple two-input gates, such as $\sqrt{\text{SWAP}}$, and does not rely on the three-input controlled-SWAP gate \cite{F02,ekertetal}, whose explicit construction is not trivial.
Moreover, the necessary ancillas, that interact with the target systems \cite{F02,ekertetal}, could be automatically supplied by projective measurements performed on a fraction of the ensemble itself of target systems. As we will show in explicit examples, our strategy does not rely on the independent preparation of ancillary systems, provided the ancillas are still available for further manipulations after the projective measurements.

This paper is organized as follows.
In Sec.\ \ref{sec:Gen}, we present the general ideas. We show here how to extract information about the purity of a quantum system.
As stated above, since purity is a quadratic functional of the density matrix, we prepare the initial state in a tensor-product form which includes two system density matrices. 
An ancilla system is introduced and coupled to the system under consideration.
A unitary operation (gate), coupling the system and the ancilla, is applied twice in succession. Finally, 
 a measurement on the ancilla yields information on the purity of the system.
Neither tomographic nor interferometric setups is introduced. 
These general ideas are then made concrete by explicitly constructing the unitary operator and by identifying the necessary measurements in the case of a two-level system (qubit) in Sec.\ \ref{sec:qubit}, a three-level system (qutrit) in Sec.\ \ref{sec:qutrit}, and a general $N$-level system in Sec.\ \ref{sec:GenN}\@.
Section \ref{sec:dis} is devoted to the summary and discussion.
An Appendix is added to describe, in terms of a  ``generalized" Bloch vector, how the scheme works in general.

\section{\bf General idea and framework}
\label{sec:Gen}
Let a quantum-mechanical system (target system) with $N$ discrete levels be described by a density matrix $\rho$, which can be characterized by a generalized $(N^2-1)$-dimensional Bloch vector $\bm a$.
Our goal is to measure its purity $\Pi=\Tr\{\rho^2\}$.
Assume that we can prepare the target state $\rho$ in duplication, i.e., $\rho\otimes\rho$.
We also prepare another quantum system (ancilla) with $N$ discrete levels in a particular pure state $\omega$.
The initial state reads 
\begin{equation}
\rho\otimes\rho\otimes\omega.
\label{eq:rho2omega}
\end{equation}
Observe that this ancilla can be supplied by a fraction of the ensemble of the target state $\rho$ by performing an appropriate projective measurement, which is required in our scheme for the estimation of purity, as we will see below, and one could thus reduce the number of different resources required.
We first make one of the duplicated target systems interact with the ancilla via a unitary evolution operator (gate) $U$.
After the interaction, the information about the target state $\rho$ has been transferred to the ancilla and they are in an entangled state.
The total system is then exposed to another interaction, this time only between the other target system and the ancilla, governed by the same type of unitary evolution operator $U$.
Finally, a suitable observable of the ancilla is measured, from which information about the purity of the target system can be extracted.

The state of the ancilla just after each step, $\omega^{(1)}$ and $\omega^{(2)}$, is expressed in the following way.
Define the map acting on two quNits  
\begin{equation}
\Lambda(\rho_1\otimes\rho_2) = \Tr_1\{U(\rho_1\otimes \rho_2)U^\dagger\},
\end{equation}
where $U$ is a unitary operator on $\mathbb{C}^N \otimes \mathbb{C}^N$ and the trace $\Tr_1$ is taken over the first system (left-hand side of the tensor product).
The states of the ancilla after the first and second step read
\begin{equation}
\omega^{(1)}=\Lambda(\rho\otimes\omega)
\label{eq:B(1)}
\end{equation}
and
\begin{equation}
\omega^{(2)} = \Lambda\bm{(}\rho\otimes\Lambda(\rho\otimes \omega)\bm{)},
\label{eq:B(2)}
\end{equation} 
respectively. 
Note that we do not measure the target systems (and therefore take the partial traces to get the ancilla state).

\begin{figure}[t]
\includegraphics[width=6.50 cm, height=3.0cm]{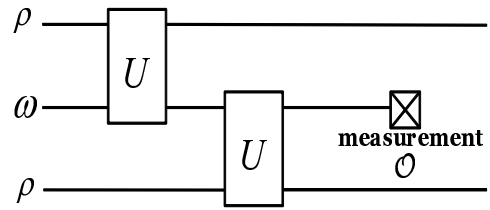}
\caption{\label{Fig1} 
Schematic diagram of the procedure. Evolution due to Eqs.\  (\ref{eq:B(1)}) and (\ref{eq:B(2)}): the output of the first gate, whose inputs are $\rho$ and $\omega$, is fed to the second gate, together with the other $\rho$. A projective measurement $\cal{O}$ is performed on the ancilla, but not on the target states (that are traced away).}
\end{figure}

Since the final state of the ancilla, represented by the reduced density matrix $\omega^{(2)}$, depends quadratically on $\rho$ and therefore quadratically on its Bloch vector $\bm a$, the expectation value of an ancilla's observable is expected to carry part of the information about ${\bm a}^2$ (because in general there remains no larger symmetry that keeps $\bm a^2$ invariant).
This is why we have presented the target state $\rho$ in duplication. 
The purity of the target state $\rho$ can now be evaluated, eventually by supplementing 
additional information about vector $\bm a$, that can be obtained by ordinary projective measurements directly on the target state $\rho$.
A schematic diagram is shown in Fig.\ \ref{Fig1}.

To be more definite, assume that the ancilla is prepared in a pure state 
\begin{equation}
\omega=|n\rangle\langle n|,
\label{eq:omega}
\end{equation}
and we measure the probability of finding the ancilla in the same state
\begin{equation}
\mathcal{O}=\openone\otimes\openone\otimes|n\rangle\langle n|.
\end{equation}
Clearly, this observation is performed on the final state $\omega^{(2)}$.
Furthermore, suppose that the interaction Hamiltonian $H$ between each target system and the ancilla is a SWAP operator
\begin{equation}
H(|i\rangle\otimes|n\rangle)\propto|n\rangle\otimes|i\rangle,\quad\forall i, n.
\end{equation}
Then it is easy to see that, since the successive gate operations induce on the basis vector  the transformation $|i,j,n\rangle\equiv|i\rangle\otimes|j\rangle\otimes|n\rangle\to|n,j,i\rangle\to|n,i,j\rangle$, the expectation value of ${\cal O}$ contains terms like 
\begin{align}
\langle{\cal O}\rangle&\sim\sum_{i,j}\langle n,i,j|(\rho\otimes\rho\otimes|n\rangle\langle n|)|i,j,n\rangle\nonumber\\
&=\sum_i\langle n|\rho|i\rangle\langle i|\rho|n\rangle \nonumber \\
&= \langle n|\rho^2|n\rangle, 
\end{align}
which are nothing but the diagonal components of $\rho^2$.
By collecting these results for a complete orthonormal set $\{|n\rangle\}$, we obtain the purity
\begin{equation}
\label{pi}
 \Pi = \Tr\{\rho^2\} = \sum_n \langle n|\rho^2|n\rangle.
 \end{equation}
This clarifies which elements are important for the above general framework to work properly:
i) a SWAP Hamiltonian, which brings about a two-body gate, and ii) a set of projective measurements on the ancilla.  
As we will see below, we shall also need iii) a direct measurement on $\rho$ to single out $\langle n|\rho^2|n\rangle$ from $\langle\mathcal{O}\rangle$.
We emphasize that the scheme requires only two simple two-body interactions (SWAPs) and does not rely on more complicated gates like the controlled-SWAP\@.
 
The general scheme described above will now be tested on two simple examples: a two-level system (qubit) and a three-level system (qutrit).
The extension to general $N$-level ($N>3$) systems is straightforward and will be presented in Sec.\ \ref{sec:GenN}\@.
Incidentally, as is clear from the above discussion, in practice one needs to calculate the reduced density matrices at each step as functions of the Bloch vector $\bm a$, or alternatively, one can directly compute the expectation value $\langle\mathcal{O}\rangle$ in the final state, as is shown in Sec.~\ref{sec:GenN}.


\section{\bf Qubit ($N=2$) case}
\label{sec:qubit}
Let a quantum-mechanical two-level system (target qubit) be described by a density matrix $\rho$, characterized by an unknown Bloch vector $\bm a$, 
\begin{equation}
\label{rhoa}
\rho=\frac{1}{2}(\openone+{\bm a}\cdot{\bm\sigma}),
\end{equation}
where $\bm{\sigma}$ is the vector of Pauli matrices.
The square modulus of $\bm a$ is related to the purity $\Pi$ of $\rho$ by
\begin{equation}
\Pi=\Tr\{\rho^2\}={1\over2}(1+{\bm a}^2),\quad
{1\over2}\le\Pi\le1.
\end{equation}
If the system is in a pure state, $\Pi=1$ and ${\bm a}^2=1$, while for the completely mixed state $\rho=\openone/2$ purity is minimal $\Pi=1/2$ and $\bm a=0$.
We introduce an ancilla qubit, prepared in a pure state 
\begin{gather}
\label{rhonpure}
\omega=|\bm n\rangle\langle\bm n|={1\over 2}(\openone+\bm n\cdot\bm\sigma), \\
(\bm n\cdot{\bm\sigma})|\bm n\rangle=|\bm n\rangle,\quad|\bm n|=1
\end{gather}
and make it interact with the target qubit. 
Assume that the unitary operator (gate) $U$ representing the interaction between each of the target qubits and the ancilla is given by
\begin{equation}
U=\frac{1}{\sqrt{2}}(\openone\otimes\openone-iS),
\label{eqn:sSWAP}
\end{equation}
which can be generated by the SWAP Hamiltonian
\begin{equation}
S=\frac{1}{2}\openone\otimes\openone+\frac{1}{2}\sum_{i=1}^3\sigma^i\otimes\sigma^i
\label{eqn:Squbit}
\end{equation} 
with the Heisenberg interaction as $U=e^{-igt S}|_{gt=\pi/4}$.
Indeed, by noting the property of the SWAP operator $S^2=\openone\otimes\openone$, one gets $U(t)=e^{-igtS}=\cos (gt) \openone\otimes\openone-iS\sin(gt)$, which reduces to (\ref{eqn:sSWAP}) at $t=\pi/4g$.
Observe incidentally that this unitary operator is just a realization of $\sqrt{\hbox{SWAP}}$, since its square yields $U^2=-iS$, which is nothing but the SWAP \cite{note:FractionalSWAP}.

It is easy to see that if $\rho_1={1\over2}(\openone+\bm{a}\cdot\bm\sigma)$ and $\rho_2={1\over2}(\openone+\bm{b}\cdot\bm\sigma)$, the map $\rho_2^{(1)}\equiv\Lambda(\rho_1\otimes \rho_2)={1\over2}(\openone+\bm c\cdot\bm\sigma)$ is mirrored into the map of the corresponding Bloch vectors
\begin{equation}
\bm{c} = \tilde{\Lambda}(\bm{a},\bm{b}) =
\frac{1}{2}(\bm{a}+\bm{b}+\bm{a}\times\bm{b}).
\label{eq:Lambdatilde}
\end{equation}
In the scheme outlined in Fig.\ \ref{Fig1}, $\rho_1$ and $\rho_2$ are given by (\ref{rhoa}), while $\omega$ is given by  (\ref{rhonpure}), so that the Bloch vector of the state of the ancilla after the first step $\omega^{(1)}={1\over2}(\openone+\bm b_1\cdot\bm\sigma)$ reads
\begin{equation}
\bm b_1=\tilde{\Lambda}(\bm{a},\bm{n})={1\over2}(\bm a+\bm n+\bm a\times\bm n).
\label{eq:a1b1}
\end{equation}
From~(\ref{eq:B(2)}), (\ref{eq:Lambdatilde}) and (\ref{eq:a1b1}), the Bloch vector $\bm b_2$ characterizing the reduced density matrix of the ancilla after the second interaction $\omega^{(2)}={1\over2}(\openone+\bm b_2\cdot\bm\sigma)$ is given by
\begin{align}
\bm b_2 &=\tilde \Lambda(\bm a,\bm b_1)= \tilde{\Lambda}\bm{(}\bm{a},\tilde{\Lambda}(\bm{a},\bm{n})\bm{)}
\nonumber\\
&= \frac{1}{4}[3\bm{a}+\bm{n}+2\bm a\times\bm n+\bm{a}\times(\bm{a}\times\bm{n})].
\end{align}
Observe that this expression is quadratic in the Bloch vector $\bm a$.
If we measure the spin of the ancilla along direction $\bm n$ in the final state $\omega^{(2)}$, its expectation value $\cal N$ is given by
\begin{align}
{\cal N}&=\Tr\{(\bm n\cdot\bm\sigma)\omega^{(2)}\}=\bm n\cdot\bm b_2\nonumber\\
&=\frac{1}{4}[1 + 3 \bm{n}\cdot\bm{a} + (\bm{n}\cdot\bm{a})^2-\bm{a}^2].
\label{nana}
\end{align}
The purity of the target state $\rho$ can thus be written as
\begin{equation}
\label{piN}
\Pi=1+{3\over2}\bm n\cdot\bm a+{1\over2}(\bm n\cdot\bm a)^2-2{\cal N}.
\end{equation}
Evaluation of this expression requires knowledge of $\bm{n}\cdot\bm{a}$, which can be easily obtained by directly measuring the spin of the target system $\rho$ along $\bm n$, 
\begin{equation}
\Tr\{(\bm n\cdot\bm\sigma)\rho\}=\bm n\cdot\bm a,
\end{equation}
or by measuring the spin of the ancilla $\omega^{(1)}$ along $\bm n$, 
\begin{equation}
\Tr\{(\bm n\cdot\bm\sigma)\omega^{(1)}\}=\bm n\cdot\bm b_1={1\over2}(1+\bm n\cdot\bm a).
\end{equation}
Both these measurements can be performed independently, e.g.\ by using a portion of the targets or ancillas.

Notice that no tomographic method involving the complete reconstruction of the state $\rho$ has been invoked; only a single type of measurement on the ancilla, that is, the measurement of the spin along direction $\bm n$, is required to obtain the purity of the target state $\rho$. In other words, there is no need to collect data with different $\bm{n}$'s.
Notice also that $\bm{n}$ can be chosen arbitrarily.

Furthermore no interferometric setup in a quantum network \cite{F02,ekertetal} is required,
significantly reducing in this way the necessary experimental steps. Indeed, only two simple $\sqrt{\text{SWAP}}$ gates are employed.

Finally, the ancilla could even be supplied from the ensemble of the target state $\rho$, 
when one performs the measurement on a part of it and finds it in the same direction $\bm n$, since the state after the measurement with an affirmative result is reduced to the state $\omega=|\bm n\rangle\langle\bm n|$.   
This also yields the necessary information, i.e.\  the quantity $\bm{n}\cdot\bm{a}$, related to the probability of finding the target system in state $\omega$.

\section{\bf Qutrit ($N=3$) case}
\label{sec:qutrit}
The extension of the above scheme to multi-level systems is straightforward (although not trivial).
In this section, we consider $N=3$ case (qutrit).  
Generalization to still higher $N$ is given in the Appendix, while a different derivation is given in the following section. 

The density matrix of a three-level system can be written as 
\begin{equation}
\rho={1\over3}\openone+{1\over\sqrt3}a^c\lambda^c\equiv{1\over3}\openone+{1\over\sqrt3}\bm a\cdot\bm\lambda,
\label{eq:rhoN}
\end{equation}
where, here and in the following, the summation over repeated indices (in this case from $1$ to $8$) is implicit, $\bm a$ is a generalized eight-dimensional Bloch vector and the generators of the SU(3) group $\lambda^a/2$ ($a=1,\ldots,8$), that are all traceless $\Tr\{\lambda^a\}=0$, satisfy the commutation and anti-commutation relations with the totally anti-symmetric and symmetric structure constants 
\begin{equation}
[\lambda^a,\lambda^b]=2if^{abc}\lambda^c,\quad
\{\lambda^a,\lambda^b\}={4\over3}\delta^{ab}+2d^{abc}\lambda^c ,
\label{eq:fandd}
\end{equation}
with the normalization condition
\begin{equation}
\Tr(\lambda^a\lambda^b)=2\delta^{ab}.
\label{eq:normalization}
\end{equation}
The purity of the system is 
\begin{equation}
\Pi=\Tr\{\rho^2\}={1\over3}+{2\over3}\bm a^2
\end{equation}
and a pure state with purity $\Pi=1$ is again characterized by a Bloch vector of unit length $|\bm a|=1$.
We have to pay due attention to the fact that not all the states expressed as in (\ref{eq:rhoN}) 
are physically acceptable.
In fact, in order to insure the positivity of the density matrix, the physical domain of the Bloch vector is much more restrictive than $|\bm a|\le1$ \cite{K03}. 
For the moment, we assume that only physically acceptable Bloch vectors have been chosen when we write the density matrix in terms of the generators $\lambda^a/2$. 
We shall return to this point later.

We now let a target qutrit, whose density matrix $\rho$ is characterized by an unknown Bloch vector $\bm a$ as in (\ref{eq:rhoN}), interact with an ancilla qutrit, prepared in a prescribed pure state
\begin{equation}
\omega={1\over3}\openone+{1\over\sqrt3}\bm n\cdot\bm\lambda,\quad
\bm n^2=1.
\end{equation}
Assume that the interaction is again $\sqrt{\text{SWAP}}$ as in (\ref{eqn:sSWAP}), with the SWAP Hamiltonian $S$ for a couple of qutrits \cite{note:SWAPqutrit}
\begin{equation}
S=\frac{1}{3}\openone\otimes\openone+\frac{1}{2}\lambda^c\otimes\lambda^c.
\end{equation}
(Additional details on this general structure can be found in the Appendix.)
After the unitary evolution engendered by this operator, the reduced density matrix of the ancilla reads 
\begin{equation}
\omega^{(1)}=\Lambda(\rho\otimes\omega)={1\over3}\openone+{1\over\sqrt3}\bm b_1\cdot\bm\lambda,
\end{equation} 
and can be shown to be characterized by the Bloch vector 
\begin{align}
\bm b_1&=\tilde\Lambda(\bm a,\bm n)={1\over2}\left(
\bm a+\bm n+{}{2\over\sqrt3}(\bm a\times\bm n)
\right),
\end{align}
where a simplified notation has been introduced for the anti-symmetric product between eight-dimensional vectors 
\begin{equation}
(\bm a\times\bm n)^a=f^{abc}a^bn^c,\quad
\bm a\times\bm n=-\bm n\times\bm a .
\end{equation}
Finally, the reduced density matrix $\omega^{(2)}$ of the ancilla after the second interaction is characterized by the Bloch vector $\bm b_2$
\begin{align}
\bm b_2&=\tilde\Lambda(\bm a,\bm b_1)=\tilde\Lambda\bm{(}\bm a,\tilde\Lambda(\bm a,\bm n)\bm{)}
\nonumber\\
&={1\over4}\bm n+{3\over4}\bm a+{1\over\sqrt3}(\bm a\times\bm n)+{1\over3}\bm a\times(\bm a\times\bm n).
\end{align}
By measuring the operator $\bm n\cdot\bm\lambda$ on the ancilla, we get the expectation value
\begin{align}
{\cal N}&=\Tr\{(\bm n\cdot\bm\lambda)\omega^{(2)}\}={2\over\sqrt3}\bm n\cdot\bm b_2\nonumber\\
&={1\over\sqrt3}\left({1\over2}+{3\over2}(\bm n\cdot\bm a)-{2\over3}(\bm a\times\bm n)^2\right).
\label{N3}
\end{align}

Notice that the last term contains a bilinear factor in $\bm a$, which can be rewritten as 
\begin{align}
&(\bm a\times\bm n)^2\nonumber\\
&\ =f^{abc}f^{cde}a^an^ba^dn^e\nonumber\\
&\ ={2\over3}[\bm a^2-(\bm n\cdot\bm a)^2]-(\bm n\star\bm a)^2+(\bm a\star\bm a)\cdot(\bm n\star\bm n),
\label{eq:acrossb}
\end{align}
where a symmetric product between two eight-dimensional vectors has been introduced, yielding another vector 
\begin{align}
(\bm a\star\bm n)^a&\equiv d^{abc}a^bn^c,\nonumber\\
\bm a\star\bm n&=\bm n\star\bm a,\ \ \bm n\cdot(\bm a\star\bm n)=\bm a\cdot(\bm n\star\bm n),\ \ {\rm etc.}
\end{align}
Recall that in dimension $D$ $(=N^2-1)$ higher than three, the magnitude of a $D$-dimensional vector $\bm a$ cannot be simply determined by its component $\bm a\cdot\bm n$ along a particular unit vector $\bm n$,  and by one of its normal components $|\bm a\times\bm n|$, since there are more than one directions normal to a plane spanned by two vectors in $D$ dimensions, which can alternatively be seen as the breakdown of the vector relation $\bm n\times(\bm a\times\bm n)=\bm a-(\bm a\cdot\bm n)\bm n$, valid in three dimensions.
This means that the above expectation value (\ref{N3}), supplemented with the value of $\bm n\cdot\bm a$, is not enough to extract the Bloch vector squared $\bm a^2$.
One may overcome this problem by measuring not a single particular operator $\bm n\cdot\bm\lambda$ fixed by a given $\bm n$, but several (in practice three) operators specified by properly chosen unit vectors.
At this point, we have to be careful in choosing $\bm n$ so that the resulting density matrix really represents a physical state \cite{K03}.
In order to trivially satisfy this physical-state condition, we choose the following unit Bloch vectors 
\begin{align}
\bm n_1&=(0,0,\sqrt3/2,0,0,0,0,1/2),\nonumber\\
\bm n_2&=(0,0,-\sqrt3/2,0,0,0,0,1/2),\nonumber\\
\bm n_3&=(0,0,0,0,0,0,0,-1),
\end{align}
which correspond to the states of the ancilla represented by the diagonal density matrices 
\begin{equation}
\begin{pmatrix}1&0&0\\0&0&0\\0&0&0\end{pmatrix},\quad\begin{pmatrix}0&0&0\\0&1&0\\0&0&0\end{pmatrix},\quad\begin{pmatrix}0&0&0\\0&0&0\\0&0&1\end{pmatrix},
\end{equation}
respectively.
Choosing one of these unit vectors simply means that we prepare the ancilla qutrit in one of the three levels and measure its population.
We easily evaluate the quadratic term in $\bm a$ by inserting the actual values of the structure constants
\begin{align}
(\bm a\times\bm n_1)^2
&={3\over4}\Bigl((a^1)^2+(a^2)^2+(a^4)^2+(a^5)^2\Bigr),\nonumber\\
(\bm a\times\bm n_2)^2
&={3\over4}\Bigl((a^1)^2+(a^2)^2+(a^6)^2+(a^7)^2\Bigr),\nonumber\\
(\bm a\times\bm n_3)^2
&={3\over4}\Bigl((a^4)^2+(a^5)^2+(a^6)^2+(a^7)^2\Bigr).
\end{align}
It is evident that these values, together with the values of $a^3$ and $a^8$, that are also available just by measuring the populations of the three levels in $\rho$ (or in $\omega^{(1)}$), are enough to determine the Bloch vector squared $\bm a^2$.
In other words, the measurements of the populations of the three levels (in states $\omega^{(2)}$ and $\rho$ or $\omega^{(1)}$) are enough to determine the purity of the qutrit system $\rho$ characterized by the (eight-dimensional) Bloch vector $\bm a$.

\section{GENERAL $N$ CASE}
\label{sec:GenN}
The qutrit case of the preceding section can be straightforwardly generalized to higher $N$, however, the results are a bit involved and are collected in the Appendix.
In this section we follow a simpler route and obtain the purity of a general $N$-level system, according to the general idea presented in Sec.~\ref{sec:Gen}.

Recall that the density matrix $\rho$ can be expressed as 
\begin{equation}
\rho=\sum_\alpha p_\alpha|\psi_\alpha\rangle\langle\psi_\alpha|,\quad\sum_\alpha p_\alpha=\Tr\{\rho\}=1,
\end{equation}
in terms of its eigenvectors $|\psi_\alpha\rangle$ belonging to the eigenvalues $p_\alpha$ ($0<p_\alpha\le1$), which are orthonormal to each other $\langle\psi_\alpha|\psi_\beta\rangle=\delta_{\alpha\beta}$.
We prepare the ancilla state $\omega$ in one of the $N$-levels, say $|n\rangle$ ($n=1,\ldots,N$), so that $\omega=|n\rangle\langle n|$.
Since our initial state is of the form (\ref{eq:rho2omega}), we first calculate the action of two successive $\sqrt{\rm SWAP}$ operations (\ref{eqn:sSWAP}) on the state $|\psi_\alpha\rangle\otimes|\psi_\beta\rangle\otimes|n\rangle\equiv|\alpha,\beta,n\rangle$, obtaining
\begin{equation}
|\alpha,\beta,n\rangle\to{1\over2}[|\alpha,\beta,n\rangle-i|\alpha,n,\beta\rangle-i|n,\beta,\alpha\rangle-|n,\alpha,\beta\rangle],
\end{equation}
where the $\sqrt{\rm SWAP}$ operation on the first and last (ancilla) entries is followed by that between the second and the last (ancilla) entries.
If the projective measurement on the state $|n\rangle$ is performed on the ancilla (i.e., the last entry), this state vector reduces to 
\begin{equation}
{1\over2}[|\alpha,\beta\rangle-i\langle n|\beta\rangle|\alpha,n\rangle-i\langle n|\alpha\rangle|n,\beta\rangle-\langle n|\beta\rangle|n,\alpha\rangle].
\end{equation}
Thus we know that the probability of finding the ancilla in state $|n\rangle$ in the final state after the two successive $\sqrt{\rm SWAP}$ operations on the initial state $\rho\otimes\rho\otimes\omega$ is simply given by
\begin{align}
&{1\over4}\sum_{\alpha,\beta}p_\alpha p_\beta\Tr\{[|\alpha,\beta\rangle-i\langle n|\beta\rangle|\alpha,n\rangle
\nonumber\\[-4truemm]
&\qquad\qquad\qquad\qquad\quad
{}-i\langle n|\alpha\rangle|n,\beta\rangle-\langle n|\beta\rangle|n,\alpha\rangle]\nonumber\\
&\qquad\qquad\qquad
\times[\langle\alpha,\beta|+i\langle n|\beta\rangle^*\langle\alpha,n|
\nonumber\\
&\qquad\qquad\qquad\qquad\qquad
{}+i\langle n|\alpha\rangle^*\langle n,\beta|-\langle n|\beta\rangle^*\langle n,\alpha|]\}\nonumber\\
&={1\over4}-{1\over2}\sum_\alpha p_\alpha^2|\langle n|\psi_\alpha\rangle|^2\nonumber\\
&\qquad+{3\over4}\sum_\alpha p_\alpha|\langle n|\psi_\alpha\rangle|^2+{1\over2}\Bigl(\sum_\alpha p_\alpha|\langle n|\psi_\alpha\rangle|^2\Bigr)^2\nonumber\\
&={1\over4}-{1\over2}{\rm Tr}\{\rho^2|n\rangle\langle n|\}\nonumber\\
&\qquad+{3\over4}{\rm Tr}\{\rho|n\rangle\langle n|\}+{1\over2}\Bigl({\rm Tr}\{\rho|n\rangle\langle n|\}\Bigr)^2.
\end{align}
Therefore, if one sums up the result for each measurement, from $n=1$ to $N$, one ends up with
\begin{equation}
{N+3\over4}-{1\over2}\Tr\{\rho^2\}+{1\over2}\sum_{n=1}^N\Bigl({\rm Tr}\{\rho|n\rangle\langle n|\}\Bigr)^2.
\end{equation}  
[In the case $N=2$ (qubit case), one easily finds that ${\rm Tr}\{\rho^2|n\rangle\langle n|\}$ contains the Bloch vector squared $\bm a^2$ for an arbitrary $|n\rangle$ and no further measurements are necessary to obtain the purity.]
Purity, represented by the second term in the above expression, is thus extracted from the result of the measurements of the ancilla state after the two $\sqrt{\rm SWAP}$ operations, if it is supplemented with the information on the population of each level ${\rm Tr}\{\rho|n\rangle\langle n\}$, which is obtained by the usual projective measurement on $\rho$.

\section{SUMMARY AND DISCUSSION}
\label{sec:dis}
The above examples explicitly show how the general idea presented in Sec.~\ref{sec:Gen} can actually be implemented in practice.
The scheme presented enables one to measure the purity of a quantum mechanical system by resorting neither to tomography nor to interferometry.
This implies that one would need a relatively small number of experimental manipulations.
Indeed, in this framework, one only needs to measure the population of each discrete level: the number of different types of measurement scales linearly with $N$, a situation to be contrasted with that encountered in the strategy that makes use of state tomography, where $N^2-1$ independent elements of the density matrix have to be determined.

As explained in Sec.~\ref{sec:Gen}, one needs to prepare the system in duplication in order to estimate its purity, since purity is a quadratic functional of the density matrix.
It is known and is easily confirmed that the expectation value of the SWAP operator $S$, which is a two-body operator, on such a duplicated state directly yields the purity of the state, $\Tr\{S(\rho\otimes\rho)\}=\Tr\{\rho^2\}$.
It is, however, not trivial to realize such multi-body measurements \cite{HDFF03}.
If one is allowed to measure only local observables, not multi-body operators, one would need one more system, such as an ancilla, to get information on the purity.
The situation is exactly what happens in our case and also in those cases where an interferometric setup is used \cite{F02,ekertetal}.

Since we need $N$ types of projective measurements for the target system and for the ancilla after the successive gate operations, and since the system has to be prepared in duplication, the number of necessary resources is $3N$, which is to be contrasted with the three resources necessary in the interferometric setup \cite{F02,ekertetal}.
(The three resources are the target system in duplication + one ancilla: the ``path'' degree of freedom in the schemes presented in \cite{F02,ekertetal} is actually made up of two levels of the ancilla qubit.)
In spite of this apparently less appealing feature, the present scheme is free from potential difficulties that are inherent in the schemes based on the controlled gates \cite{F02,ekertetal}.

The ideas put forward in \cite{F02,ekertetal} were implemented in two experiments.
In Ref.~\cite{HDFF03}, a SWAP (flip) operator was directly measured to obtain the overlap of photonic polarization states, without introducing an interferometric setup, while in \cite{X03} the fidelity of two NMR qubits was measured. 
The scheme discussed in this article, on the other hand, makes use of two-body unitary gates, that can be easily realized by means of a familiar interaction Hamiltonian. Moreover, no controlled gate is necessary. 
The projective measurement at each level, required in the present scheme, is nothing but a measurement of the population of that level.

Finally, the examples presented in this paper depend on the practical realization of SWAP Hamiltonians and this may not be a trivial problem for general $N$-level systems, though the interaction reduces to the familiar spin-exchange one in the qubit ($N=2$) case.  
In this respect, it is an open problem, yet to be clarified, whether the present scheme could be generalized to other Hamiltonians.
On the other hand, the general ideas presented  in Sec.~\ref{sec:Gen} could be extended to more general nonlinear situations, e.g.\ in order to estimate $\Tr\{\rho^n\}$ with $n>2$.
It would be worth exploring such a possibility along the same line of thought, since the scheme only requires simple two-body gate operations and projective measurements and is thus relatively easy to implement in experiments.

\acknowledgments
The authors acknowledge inspiring discussions with Paolo Facchi and Vittorio Giovannetti, that have greatly influenced their work. 
This work is partly supported by a Grant-in-Aid for Scientific Research (C) from JSPS, Japan and by the Joint Italian-Japanese Laboratory on ``Quantum Technologies: Information, Communication and Computation" of the Italian Ministry of Foreign Affairs.
K.Y. is supported by the Program to Disseminate Tenure Tracking System and the Grant-in-Aid for Young Scientists (B) (No.\ 21740294) both from the Ministry of Education, Culture, Sports, Science and Technology, Japan.

\appendix*
\section{General $N$ case: Another exposition}
\label{app:GenN}

We rederive here the result obtained in Sec.\ \ref{sec:GenN} by using an alternative method, with the hope that the different mathematical techniques employed will aid in future exploration of the field. The derivation is more lenghty, but can lead to generalization for higher-order functionals of the density operator. 

The density matrix of an $N$-level system can be written as
\begin{equation}
\label{rhoN}
\rho={1\over N}\openone+{\sqrt{\frac{N-1}{2N}}}a^cT^c\equiv{1\over N}\openone+{\sqrt{\frac{N-1}{2N}}}\bm a\cdot\bm T,
\end{equation}
where the index $c$ runs from $1$ to $N^2-1$, $\bm a$ is an $(N^2-1)$-dimensional Bloch vector, and the generators of the SU($N$) group $T^a/2$, that are traceless $\Tr\{T^a\}=0$, satisfy
\begin{equation}
[T^a,T^b]=2if^{abc}T^c, \quad
\{T^a,T^b\}={4\over N}\delta^{ab}+2d^{abc}T^c
\end{equation}
with the normalization condition
\begin{equation}
\Tr(T^aT^b)=2\delta^{ab}.
\end{equation}
Here $f^{abc}$ and $d^{abc}$ are the SU($N$) totally anti-symmetric and symmetric structure constants.
One can also derive the relations
\begin{gather}
T^aT^b={2\over N}\delta^{ab}+(d^{abc}+if^{abc})T^c,\\
T^a_{ij}T^a_{k\ell}=-{2\over N}\delta_{ij}\delta_{k\ell}+2\delta_{i\ell}\delta_{jk}.
\end{gather}
The structure constants are written in terms of the generators as
\begin{equation}
d^{abc}={1\over4}\Tr(T^a\{T^b,T^c\}),\quad
f^{abc}={1\over4i}\Tr(T^a[T^b,T^c]),
\end{equation}
from which the following relations follow
\begin{gather}
f^{abc}f^{abd}=N\delta^{cd},\quad
d^{abc}d^{abd}={N^2-4\over N}\delta^{cd},\quad
d^{aab}=0,\nonumber\\
f^{abc}f^{cde}={2\over N}(\delta^{ad}\delta^{be}-\delta^{bd}\delta^{ae})+d^{adc}d^{ceb}-d^{bdc}d^{cea}.
\end{gather}
The purity of the system in terms of the Bloch vector $\bm a$ reads 
\begin{equation}
\Pi=\Tr\{\rho^2\}={1\over N}+{N-1\over N}\bm a^2,
\label{eq:PiN}
\end{equation}
and thus a pure state with purity $\Pi=1$ is characterized by a Bloch vector of unit length $\bm a^2=1$.

A quantum $N$-level system, described by the density matrix (\ref{rhoN}), with $\bm a$ unknown,
is coupled with another quantum $N$-level system (ancilla), prepared in a pure state
\begin{equation}
\omega={1\over N}\openone+{\sqrt{\frac{N-1}{2N}}}\bm n\cdot\bm T,\quad
\bm n^2=1,
\end{equation}
by the $\sqrt{\text{SWAP}}$ gate $U$ as in (\ref{eqn:sSWAP}) with the SWAP Hamiltonian $S$ for a couple of $N$-level systems
\begin{equation}
S=\frac{1}{N}\openone\otimes\openone+\frac{1}{2}T^c\otimes T^c.
\end{equation}
After this unitary gate, the Bloch vector $\bm b_1$ characterizing the reduced density matrix $\omega^{(1)}$ of the ancilla is
\begin{equation}
\bm b_1={1\over2}(\bm a+\bm n)+\sqrt{\frac{N-1}{2N}}\bm a\times\bm n,
\end{equation}
where the same notation as in the text has been used for the anti-symmetric product between $(N^2-1)$-dimensional vectors $(\bm a\times\bm n)^a=f^{abc}a^bn^c$.
After the second gate, again given by the unitary operator (\ref{eqn:sSWAP}), acting on the state $\rho$ of another target system and the state $\omega^{(1)}$ of the ancilla characterized by $\bm b_1$ (see Fig.\ \ref{Fig1}),
the Bloch vector characterizing the reduced density matrix $\omega^{(2)}$ reads 
\begin{align}
\bm b_2&=\tilde\Lambda(\bm a, \bm b_1)=\tilde\Lambda\bm{(}\bm a,\tilde\Lambda(\bm a,\bm n)\bm{)}\nonumber\\
&={1\over4}\bm n+{3\over4}\bm a-\sqrt{N-1\over2N}\bm n\times\bm a-{N-1\over2N}\bm a\times(\bm n\times\bm a).
\end{align}
We measure the operator $\bm n\cdot\bm T$ of the ancilla in its final state $\omega^{(2)}$ and get its expectation value
\begin{align}
{\cal N}&=\Tr\{(\bm n\cdot\bm T)\omega^{(2)}\}=\sqrt{\frac{2(N-1)}{N}}\bm n\cdot\bm b_2\nonumber\\
&=\sqrt{2(N-1)\over N}\left({1\over4}+{3\over4}\bm n\cdot\bm a-{N-1\over2N}(\bm a\times \bm n)^2\right).
\label{ntrace}
\end{align}

In order to extract $\bm a^2=\sum_{k=1}^{N^2-1}(a^k)^2$, we measure not the single operator $\bm n\cdot\bm T$ of the ancilla, but the $N-1$ operators specified by properly chosen unit vectors, corresponding to $N$ diagonal density matrices
\begin{equation}
\begin{pmatrix}
1&&&\\&0&&\\&&\ddots&\\&&&0
\end{pmatrix},\ldots,
\begin{pmatrix}
0&&&\\&\ddots&&\\&&0&\\&&&1
\end{pmatrix}.
\label{eq:diagonalN}
\end{equation}
These unit vectors can certainly be written as linear combinations of $\tilde{\bm n}_3,\tilde{\bm n}_8,\ldots,\tilde{\bm n}_{(N-1)^2 -1}$, and $\tilde{\bm n}_{N^2 -1}$, where $\tilde{\bm n}_\alpha$ ($\alpha=k^2-1;\,k=2,\ldots,N$) is the unit vector pointing in the direction of the $(k-1)$th Casimir operator, i.e., $T^\alpha=\tilde{\bm n}_\alpha\cdot\bm T$.
Choosing one of such unit vectors, corresponding to one of the diagonal density matrices in (\ref{eq:diagonalN}), means to prepare the ancilla in a pure state in which only one of the $N$ levels is populated and to measure its population after the successive gate operations.
In this case, the term quadratic in $\bm a$ in (\ref{ntrace}) contains terms like $(\bm a\times\tilde{\bm n}_\alpha)\cdot(\bm a\times\tilde{\bm n}_\beta)$, where $\alpha,\beta\in\{3,8,\ldots,N^2-1\}$.
Observe that
\begin{equation}
\label{summation}
(\bm a\times\tilde{\bm n}_\alpha)\cdot(\bm a\times\tilde{\bm n}_\beta)
={1\over8}(\bm a\cdot\bm T)_{ij}(\bm a\cdot\bm T)_{ji}(T^\alpha_{ii}-T^\alpha_{jj})(T^\beta_{ii}-T^\beta_{jj}),
\end{equation}
where the summations over $i$ and $j$ are understood and the fact that both Casimir operators $T^\alpha$ and $T^\beta$ are diagonal matrices has been used.

We now show that the summation of (\ref{summation}) yields a desired result, eliminating unwanted terms like the last two terms on the right hand side of (\ref{eq:acrossb}), and leaving only relevant terms.
Let $\bm n_i$ ($i=1,\ldots,N)$ be the unit vector corresponding to the $i$th-level pure state whose density matrix is of the above diagonal form (\ref{eq:diagonalN}) with the only matrix element 1 at the $(i,i)$ component.
It is expressed as a linear combination of $\tilde{\bm n}_\alpha$
\begin{equation}
\bm n_i=\gamma^{(i)}_\alpha\tilde{\bm n}_\alpha
\end{equation}
and the density matrix reads 
\begin{equation}
|i\rangle\langle i|={1\over N}\openone+\sqrt{N-1\over2N}\bm n_i\cdot\bm T={1\over N}\openone+\sqrt{N-1\over2N}\gamma^{(i)}_\alpha T^\alpha.
\end{equation}
Here the repeated Greek indices mean the summation over the diagonal (Casimir) operators $\alpha\in\{3,\ldots,$$N^2-1\}$. 
Since the coefficient $\gamma^{(i)}_\alpha$ is given by
\begin{equation}
\gamma^{(i)}_\alpha={1\over2}\sqrt{2N\over N-1}T^\alpha_{ii}
\end{equation}
(with no summation over $i$ on the right hand side), one obtains
\begin{equation}
\sum_{i=1}^N\gamma^{(i)}_\alpha\gamma^{(i)}_\beta={N\over2(N-1)}\sum_iT^\alpha_{ii}T^\beta_{ii}={N\over N-1}\delta_{\alpha\beta}.
\end{equation}
If one prepares the ancilla in level $i$, measures its population after the gate operations and then sums up
the results for all the measurements from $i=1$ to $N$, the term quadratic in $\bm a$ reads (with summations made explicit here)
\begin{align} 
&\sum_{i=1}^N(\bm a\times\bm n_i)\cdot(\bm a\times\bm n_i)\nonumber\\
&\quad
=\sum_{i\alpha\beta}\gamma^{(i)}_\alpha\gamma^{(i)}_\beta(\bm a\times\tilde{\bm n}_\alpha)\cdot(\bm a\times\tilde{\bm n}_\beta)\nonumber\\
&\quad
={N\over8(N-1)}\sum_{\alpha\ell m}(\bm a\cdot\bm T)_{\ell m}(\bm a\cdot\bm T)_{m\ell}\nonumber\\
&\qquad\qquad\qquad\quad{}\times(T^\alpha_{\ell\ell}T^\alpha_{\ell\ell}+T^\alpha_{mm}T^\alpha_{mm}-2T^\alpha_{\ell\ell}T^\alpha_{mm})\nonumber\\
&\quad
={N\over2(N-1)}\Bigl(\Tr\{(\bm a\cdot\bm T)^2\}-\sum_\ell(\bm a\cdot\bm T)_{\ell\ell}(\bm a\cdot\bm T)_{\ell\ell}\Bigr)
\nonumber\displaybreak[0]\\
&\quad
={N\over N-1}\sum_{k\not=\alpha}(a^k)^2={N\over N-1}\Bigl(\bm a^2-\sum_\alpha(a^\alpha)^2\Bigr).
\end{align}
It is evident that together with the information on $a^\alpha$, that can be obtained by measuring the population of each level in $\rho$, $\Tr\{(\bm n_i\cdot\bm T)\rho\}$, as 
\begin{align}
a^\alpha&=\tilde{\bm n}_\alpha\cdot\bm a
={N-1\over N}\sum_{i=1}^N\gamma^{(i)}_\alpha(\bm n_i\cdot\bm a)\nonumber\\
&={1\over2}\sum_{i=1}^NT^\alpha_{ii}\Tr\{(\bm n_i\cdot\bm T)\rho\},
\end{align}
one is able to estimate $\bm a^2$ and thus purity $\Pi$ via (\ref{eq:PiN}) from the experimental data $\Tr\{(\bm n_i\cdot\bm T)\omega^{(2)}\}$ and $\Tr\{(\bm n_i\cdot\bm T)\rho\}$ $(i=1,\ldots,N)$ (the populations of each level in $\omega^{(2)}$ and $\rho$).

It is interesting to note that for an $N$-level system purity is expressed as a function of invariants (Casimirs) of SU($N$). This is a peculiarity of the method proposed ($\sqrt{\text{SWAP}}$ gates in succession, as in Fig.\ \ref{Fig1}), but might be of more general significance. We leave this issue for future investigation.


\begin{thebibliography}{99}
\bibitem{note:FractionalSWAP}
It is also clear how to realize a $(\text{SWAP})^r$ gate with a fractional power $0<r<2$.
\bibitem{note:SWAPqutrit}
This interaction is a direct generalization of the spin-spin interaction in the qubit case and is expressed in terms of the three basis vectors $|+\rangle\equiv(1,0,0),\,|0\rangle\equiv(0,1,0),\,|-\rangle\equiv(0,0,1)$, as 
$\lambda^c\otimes\lambda^c=2(\sum_{k=+,0,-}|k,k\rangle\langle k,k|+|+,0\rangle\langle0,+|+|0,-\rangle\langle-,0|+|+,-\rangle\langle-,+|+|0,+\rangle\langle+,0|+|-,0\rangle\langle0,-|+|-,+\rangle\langle+,-|)-2/3=2S-2/3$,
where $|k,k'\rangle\equiv|k\rangle\otimes|k'\rangle$, $\langle k,k'|\equiv\langle k|\otimes\langle k'|$, and $S$ is the SWAP operator.
\end{thebibliography}
\end{document}